# Beware the Gini Index! A New Inequality Measure

Sabiou Inoua

Chapman University

inoua@chapman.edu

Abstract. The Gini index underestimates inequality for heavy-tailed distributions: for example, a Pareto distribution with exponent 1.5 (which has infinite variance) has the same Gini index as any exponential distribution (a mere 0.5). This is because the Gini index is relatively robust to extreme observations: while a statistic's robustness to extremes is desirable for data potentially distorted by outliers, it is misleading for heavy-tailed distributions, which inherently exhibit extremes. We propose an alternative inequality index: the variance normalized by the second moment. This ratio is more stable (hence more reliable) for large samples from an infinite-variance distribution than the Gini index paradoxically. Moreover, the new index satisfies the normative axioms of inequality measurement; notably, it is decomposable into inequality within and between subgroups, unlike the Gini index.

# 1 Overview

The Gini index, proposed by C. Gini [1], is the most popular inequality measure. Iconic by its geometric interpretation in terms of the Lorenz curve [2], the Gini index is also fascinating by its rich mathematical properties and alternative formulations [3]. Yet the Gini index has some limitations; for example, it does not apply to a zero-mean distribution (such as the normal distribution) and it may behave poorly, falling outside the interval [0,1], for variables assuming negative values. This problem is easily fixed however, for one can adjust the definition of the index for negative values [4]. In fact, the Gini ratio's denominator is more rigorously defined for a signed variable to be the variable's mean absolute value (rather than the raw mean): then the ratio is well-defined and well-behaved for any nonzero variable (Section 4).

More importantly, the Gini index underestimates inequality for heavy-tailed distributions.[1] Thus a Pareto distribution with an exponent of 1.5 (an infinite-variance variable) has the same Gini index as any exponential distribution (a mere 0.5), because the Gini index is relatively insensitive to a distribution's tail, namely to extreme realizations of a variable (Section 2). While a statistic's robustness to extremes is desirable for data potentially distorted by outliers, it is misleading for heavy-tailed distributions, which inherently exhibit extremes. We propose an alternative inequality index (Section 3) that may seem paradoxical at first sight but that is well-behaved upon scrutiny: the variance normalized by the second moment. This ratio is stable for large samples even when the theoretical (population) variance is infinite: it is then more stable statistically than the Gini index, surprisingly (Section 4). Moreover, the new index satisfies the normative axioms of inequality measurement: notably it is decomposable into inequality within and between subgroups, unlike the Gini index.

---

[1] This limitation of the Gini index should not be confused with other seemingly related issues suggested in the literature. For example, the Gini index is commonly reported to be over-sensitive to changes in the middle of a distribution compared to changes in the distribution's tails, a claim that does not stand closer scrutiny, however [5]. This paper focuses not so much on the Gini's different treatments of portions of a distribution than the index's mild treatment of heavy tails. Other reported flaws of the index include a small-sample bias and a bias due to grouping [6, 7, 8, and refs. therein].

## 2 The Gini Index and Heavy Tails

The Gini index of a non-negative variable $X$ with positive mean is by definition:[2]

$$G(X) = \frac{\frac{1}{2}\mathbb{E}|X - X'|}{\mathbb{E}(X)}, \qquad (1)$$

where $X'$ is an independent copy of $X$ (that is, $X$ and $X'$ are independent but identically distributed). For a continuous variable $X$, the Gini index is more easily computed from the distribution function $F(x) = \text{prob}\{X \leq x\}$ through the formula:

$$G(X) = \frac{\int_0^\infty F(x)[1 - F(x)]dx}{\mathbb{E}(X)}. \qquad (2)$$

For an empirical variable (or sample) $X = [x_1, ..., x_N]$, the Gini index becomes

$$G(X) = \frac{\frac{1}{2N^2}\sum_{i=1}^{N}\sum_{j=1}^{N}|x_i - x_j|}{\text{mean}(X)}. \qquad (3)$$

This formula becomes much simpler if one arranges $X$ in ascending order, $x_{(1)} \leq x_{(2)} \leq ... \leq x_{(i)} \leq ... \leq x_{(N)}$. Then

$$G(X) = \frac{2}{N}\frac{\sum_{i=1}^{N}ix_{(i)}}{\sum_{i=1}^{N}x_i} - \frac{N+1}{N}. \qquad (4)$$

These are standard facts about the Gini index that we remind for the sequel.[3]

The following Proposition (proven shortly) illustrates the above-mentioned inadequacy of the Gini index for heavy-tailed distributions:

---

[2] Throughout this paper, the word "variable" is used preferably to "random variable". We always exclude throughout this paper the trivial case where $X=0$ with probability 1.
[3] For a review and proof of these facts and many others, see e.g. [3].

***Proposition 1.*** *If Z is an exponential distribution with mean 2/3, then Z and exp(Z) have the same Gini index G = 0.5. Moreover, G(Z) ≥ G[exp(Z)] if mean(Z) ≥ 2/3.*

This is not a minor anomaly, for Z and exp(Z) are not different just quantitatively (though this is important enough in itself): exp(Z) is a heavy-tailed variable with infinite-variance: a Pareto distribution with exponent $\alpha = 3/2$ (Figure 1).

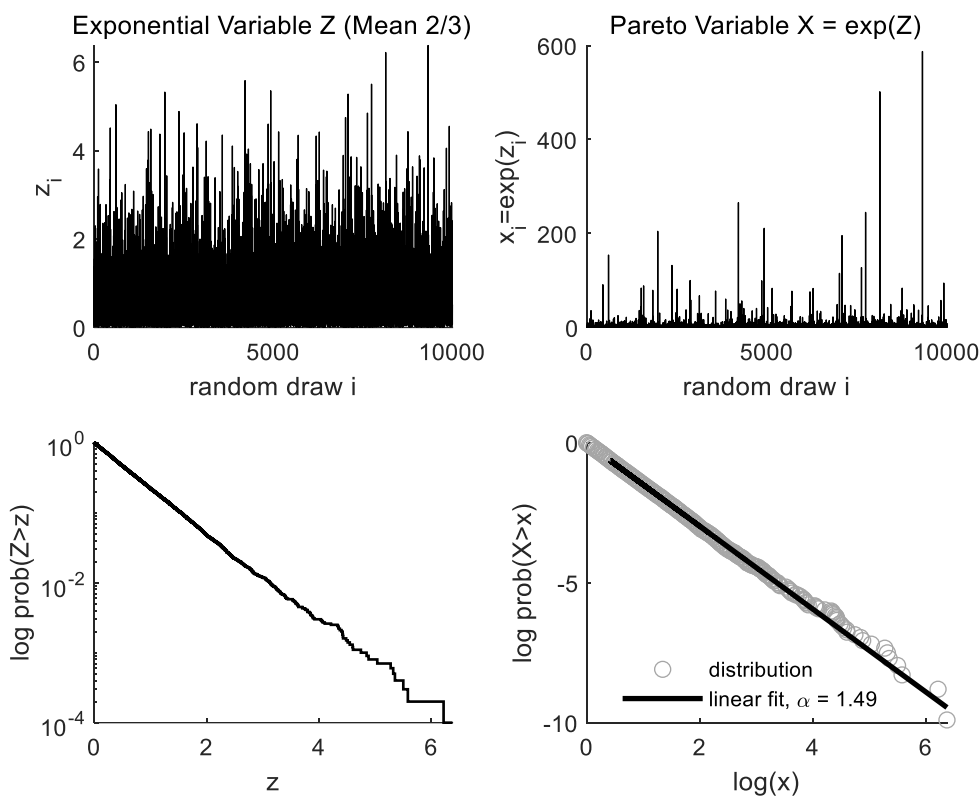

Figure 1. Exponential versus Pareto distribution. The two distributions, Z and X=exp(Z), have the same Gini index G=0.5!

A brief overview on heavy-tailed distributions here may help better articulate the sequel. A heavy-tailed variable X is precisely one that has heavier tail relatively to any exponential distribution, in the sense that $\mathbb{E}[\exp(X^\mu)] = \infty$ for all $\mu > 0$. In most applications, it is sufficient to restrict to the subclass of sub-exponential heavy-tailed variables, whose defining property is to exhibit the "one-big jump phenomenon", whereby the distribution is dominated by one observation (the maximum one), in the sense that for independent copies $\{X_1,...,X_N\}$ of a sub-exponential $X$ we have (by definition):[4]

$$\frac{\text{prob}\{X_1+...+X_N > x\}}{\text{prob}\{\max(X_1,...,X_N) > x\}} \to 1 \ (x \to \infty, N \geq 1). \tag{5}$$

An implication of this definition is that the tail of $X$ decays more slowly indeed than any exponential distribution in the sense that

$$\frac{\text{prob}\{X > x\}}{\exp(-\lambda x)} \to \infty \ (x \to \infty, \text{ all } \lambda > 0). \tag{6}$$

Exponential distributions, therefore, are a reference middle between heavy and light tails. Pareto distributions are the simplest type of heavy-tailed variables. A Pareto distribution, to remind the basics, is a continuous variable $X$ with[5]

$$\text{prob}\{X \geq x\} = (x/x_{\min})^{-\alpha}, \ x \geq x_{\min} > 0, \alpha > 0 \tag{7}$$

where the parameter $\alpha$ is called the tail exponent (or Pareto index). A Pareto distribution is therefore a line in log-log plot: $\log \text{prob}\{X \geq x\} = -\alpha \log(x) + \alpha \log x_{\min}$ (Figure 2: 4th Subplot). It can be characterized as the exponential of an exponential distribution: for any $Z$ with $\text{prob}\{Z \geq z\} = \exp[-\alpha(z - z_{\min})]$, $z \geq z_{\min}$, $\text{prob}\{e^Z \geq x\} = \text{prob}\{Z \geq \log x\} = \exp[-\alpha(\log x - z_{\min})] = (x/x_{\min})^{-\alpha}$, where $x_{\min} = \exp(z_{\min})$; conversely, $\log X$ is exponential if $X$ is Pareto. A power law $X$ is a distribution that is Pareto asymptotically, in the sense that there is $\alpha > 0$ such that $x^\alpha \text{prob}\{|X| \geq x\} \to C > 0$ as $x \to \infty$. Power laws are an important class of heavy-tailed distributions due to their ubiquity in social phenomena, notably in economic data [12, 13].

Heavy-tailed variables are special because they need not obey the standard laws of large numbers when their variability is too high (the lower $\alpha$, the more extreme the variability of a power law); and when they do, their sum converges more slowly to a

---

[4] For a review of subexponential variables, see [9-11].
[5] For a more detailed introduction to the theory and empirics of Pareto distributions, see, e.g., [12].

Gaussian distribution compared to more common distributions. One can show that for any power law, $\mathbb{E}(X^\mu) = \infty$ if $\mu \geq \alpha$, which follows by direct integration for a Pareto, for which $\mathbb{E}(X^\mu) = \alpha x_{min}^\mu / (\alpha - \mu)$ if $0 < \mu < \alpha$. Thanks to seminal work by P. Levy [14] and the contributions of other influential probability theorists of the past century, the central limit theorem has been extended to any variable (with light or heavy tails) into what we may call the "stable limit theorem", a complete characterization of all the possible limit laws of the (properly normalized) sum of independent copies of a variable. (See this paper's Appendix for a specialization of this theorem invoked throughout this paper, notably in Proposition 4, Section 4.) The only possible limit laws are called alpha-stable or Levy variables, in reference to their stability by convolution (that is, up to standardization, a stable law is preserved by addition of independent copies of it) and their characteristic parameter $\alpha \in (0,2]$: except the Gaussian (the finite variance case, or $\alpha = 2$), Levy variables are a special subclass of infinite-variance power laws whose exponents $\alpha \in (0,2)$.[6] Thus the limit sum of independent copies of any infinite-variance variable $X$, if this limit exists, is necessarily an infinite-variance variable itself, a Levy with $\alpha \in (0,2)$, unless the tail of $X$ is not heavy enough, in the sense that $x \mapsto \mathbb{E}(X^2 1_{|X| \leq x})$ is a slowly varying function: this last condition is the general condition for convergence to a Gaussian (generalizing the finite second moment condition).[7] A function $L$ is slowly varying if it behaves asymptotically like a constant, in the sense that $L(tx)/L(x) \to 1$ as $x \to \infty$ for any $t > 0$ (two common examples being a constant and a logarithmic function).[8] A necessary condition for the sum of independent copies of a variable $X$ to converge to a stable law is that $X$ have a regularly varying distribution, meaning prob{|X|

---

[6] For a synthetic textbook exposition of the general central limit theorem see, e.g., [15, sec. 2.2, th. 2.2.15] and [16, sec. 3.8, th. 3.8.2]. For intuitive derivations in a physics context, see e.g. [17, 18]. For an economic application, see e.g. [19].

[7] Throughout, $1_A = 1$ stands for the indicator function of $A$: $1_A = 1$ if $A$ is true, $1_A = 1$ if A is false.

[8] A constant function $C$ satisfies $C(xt) = C(x)$.

$\geq x\} = L(x)x^{-\alpha}$ with $L$ slowly varying: power laws are the subclass with $L(x) \to C > 0$. We focus in this paper mostly on power laws.

Going back to Proposition 1, the Gini index of a Pareto distribution with exponent $\alpha > 1$ is $G = 1/(2\alpha - 1)$, hence $G \leq 0.5$ if $\alpha \geq 3/2$ with equality if $\alpha = 3/2$; the Gini index of any exponential distribution is 0.5. [These are known facts: one can derive them from formula (2), for example.][9] This proves Proposition 1.

The Gini index, therefore, understates extreme variability. Yet the economic data of interest (such as income or wealth distributions) are precisely of this type, as is known since V. Pareto's discovery of this class of distributions [20].

The cause of this drawback of the Gini index is due to the nature of the dispersion concept underlying it, namely its numerator also known as the Gini mean difference (GMD), whose empirical (or sample) version is:[10]

$$GMD(X) = \frac{\sum_{i=1}^{N}\sum_{j=1}^{N}|x_i - x_j|}{2N^2}. \tag{8}$$

The GMD being the numerator in formula (4), it also reduces to

$$GMD(X) = \frac{2}{N^2}\sum_{i=1}^{N} i x_{(i)} - \frac{N+1}{N}\text{mean}(X). \tag{9}$$

The sensitivity of the GMD to a rank-preserving variation of a particular observation (holding the others constant) follows from (9) by simple differentiation:

$$\frac{\partial GMD(X)}{\partial x_{(k)}} = \frac{2}{N^2}[k - \frac{N+1}{2}]. \tag{10}$$

Thus, the sensitivity of the Gini index to an observation depends merely on the observation's rank, and a similar conclusion holds for the Gini ratio.[11] Compare the sensitivity of the variance, which depends on the size of the observation:

---

[9] For a shifted exponential $X = x_{\min} + Z$, $G < 0.5$ as is easily seen from (1): $G(x_{\min} + Z) < G(Z)$ because the change $Z \to Z + x_{\min}$ leaves the Gini numerator unchanged while the denominator, the mean, increases by $z_{\min}$; thus, Proposition 1 holds only for an ordinary exponential $Z$ with $z_{\min} = 0$.
[10] Some authors define the GMD as twice the numerator of the Gini ratio. The properties of the GMD are reviewed e.g. in [21].
[11] For a more detailed such sensitivity calculation, see [5].

$$\frac{\partial \operatorname{var}(X)}{\partial x_i} = \frac{2}{N}[x_i - \operatorname{mean}(X)]. \tag{11}$$

Because empirical data are possibly contaminated by a few outliers, one usually assumes (for good reasons) that the lower a statistic's sensitivity to an individual observation, the better: hence a certain advocacy for robust dispersion measures like the mean absolute deviation (MAD), or the GMD, compared to the variance.[12] Yet it is a mistake to transpose the outlier argument to a heavy-tailed distribution, which intrinsically exibits extremes that, moreover, represent a sizable portion of the distribution by definition [equation (5)]. For power laws (or regularly varying distributions more generally), the one-big jump phenomenon presents itself in a specific manner that is the more striking, the smaller the power law's tail exponent $\alpha$: in particular, when $\alpha < 1$ one can show that the maximum observation in a large random sample from a (positive) power law is on average of the same order of magnitude as the sum of all observations [22, p. 465]:[13]

$$\mathbb{E}(\frac{X_1 + ... + X_N}{\max\{X_1,...,X_N\}}) \to \frac{1}{1-\alpha} \quad (0 < \alpha < 1, N \to \infty). \tag{12}$$

If $\alpha = 0.5$, for example, one single observation in a sufficiently large sample will represent on average 50% of the sum of all observations (Figure 2).

---

[12] The MAD is even less sensitive to extremes in the sense that its sensitivity to an observation depends merely on the sign of this observation's distance from the median.

[13] For an intuitive derivation of this formula, and the corresponding ones for $\alpha > 1$, see [18, p. 106].

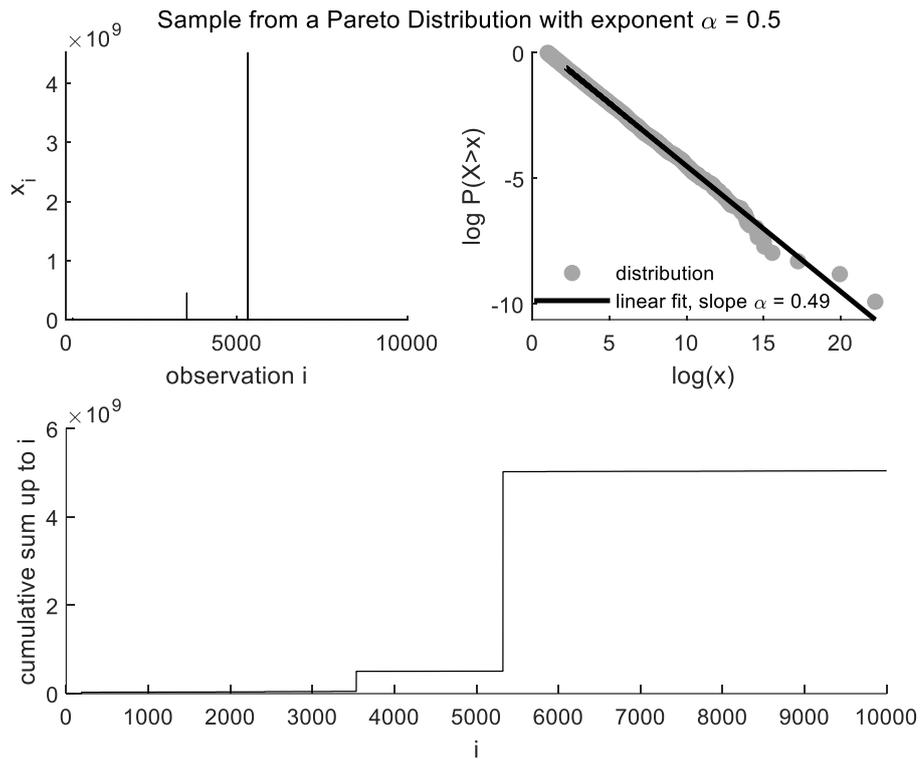

Figure 2. The one-big jump phenomenon (or dominance of a few observations) is an intrinsic feature of heavy-tailed distribution: here, a Pareto distribution with exponent 0.5; one observation (the 5324th one) accounts for much of the sum of the distribution.

Thus, any robust summary of a heavy-tailed variable is a poor summary of that distribution.

By its greater sensitivity to extremes, the variance is a more faithful measure of dispersion for power laws than the GMD. That the variance is infinite for a broader class of distributions can itself be counted as a virtue in this respect.[14] The concept of infinite variance should not deter us, for the difficulty is not as important as it may seem at first sight: an infinite theoretical (population) variance empirically means that the sample variance fluctuate wildly from sample to sample (which is just another way of saying that the distribution is ill-represented by an average value): formally, the second moment of a large sample from an infinite-variance power law is an infinite variance Levy variable by the general central limit theorem.[15] But if the variance is properly normalized into a ratio, through a denominator whose sample fluctuation is of the same order of magnitude as that of the variance, then it yields a statistically stable measure of variability (Section 4). The right denominator to that effect is none other than the second moment: as it turns out, the variance normalized by the second moment is a natural inequality measure.

---

[14] The GMD of a distribution is finite whenever the mean of the distribution is finite; thus, the GMD of a power law with exponent $\alpha > 1$ is finite. More on this in Section 4.

## 3 An Alternative Inequality Measure

We propose, therefore, for any (nonzero) empirical variable $X = [x_1,...,x_N]$, the inequality measure:

$$I(X) = \frac{\text{variance}(X)}{\text{mean}(X^2)}. \tag{13}$$

Clearly, $0 \leq I \leq 1$. Besides the obvious interpretation as a normalized variance, the ratio $I$ has various other interpretations that illustrate its potential versatility as a general concept (beyond the specific use as an inequality measure). Suffice it here to mention a few known results across the sciences that implicitly involve the inequality index $I$ and on which therefore new light may be shed if this connection is made explicit. Consider, to begin with, the following simple but powerful result in probability theory (usually known as the "second moment method"):

***Second moment method.*** *For any finite-variance $X \geq 0$ we have* $\text{prob}\{X = 0\} \leq I(X)$.

This result is a consequence of Cauchy-Schwarz inequality, it is known:

$$\mathbb{E}(X) = \mathbb{E}(X1_{\{X>0\}}) \leq [\mathbb{E}(X^2)]^{1/2}[\mathbb{E}(1_{\{X>0\}})^2]^{1/2} = [\mathbb{E}(X^2)]^{1/2}[\text{prob}\{X > 0\}]^{1/2},$$

from which (14) follows, if one solves for $\text{prob}\{X > 0\}$.

---

[15] More precisely, if $\{X_1,...,X_N\}$ are independent copies of an (infinite-variance) power law with $\alpha \in ]0,2[$, then $N^{-2/\alpha} \sum_{i=1}^{N} x_i^2$ converges in distribution to Levy variable with index $\alpha/2$. See Appendix.

The *I* index has also a series of potential applications, notably in economics, inherited from its simple relationship with the Hirshman-Herfindal index [23, 24]:

$$H(X) = \frac{\sum_{i=1}^{N} x_i^2}{(\sum_{i=1}^{N} x_i)^2}. \tag{15}$$

Clearly, we have

$$I(X) = 1 - \frac{1}{NH(X)}. \tag{16}$$

Thus, any interpretation of the *H* index can be rephrased in terms of the *I* index. The *H* index is of course primarily used as a concentration measure: *H* is then interpreted as measuring an inverse effective number of observations in a distribution (the number of dominant elements in it so to speak). That is, a variable *X* is mostly concentrated on $1/H(X)$ number of its elements (e.g. $1/H(X) = 2.05$ for $X = [1,2,1,90,110]$); in particular, perfect equality ($X = [x,…,x]$) means $1/H(X) = N$, and perfect inequality is approached by a large distribution concentrated on one individual, for which $1/H(X) = 1 - 1/N$. It makes sense therefore to measure the degree of equality in a distribution *X* by the fraction of its dominant elements, namely by the ratio $[1/H(X)]/N = 1 - I(X)$, or the effective number of elements in *X* relatively to that of a perfectly equal distribution of the same size.

A similar interpretation of *I* follows from the following probabilistic interpretation of *H*. To any nonnegative variable $X = [x_1,…,x_N]$ we associate the two probability systems $P = [p_i]$ and $Q = [s_i]$, where $p_i = 1/N$ and $q_i = x_i / \sum_{k}^{N} x_k$ ($i = 1,…, N$), and denote by $\mathbb{E}_P$ and $\mathbb{E}_Q$ the corresponding expectation operators. The probability systems can be called the uniform versus "rich-get-richer" probability measures (as would suggest the following experiment: pick at random a value from *X*, where the probability of picking $x_i$ is $p_i$, and increase the picked value by some amount; consider the same experiment, but with probability $q_i$ instead.) We have $\mathbb{E}_P(P) =$ $1/N$ and $\mathbb{E}_Q(Q) = H$. Thus, $1 - I(X)$ measures the degree of equality in $X$ by comparing the two associated probability measures through the ratio $\mathbb{E}_P(P)/\mathbb{E}_Q(Q)$.

The index $I$ also has an informational interpretation: since the inverse ratio $1/H$ measures the effective number of elements in a distribution, it follows that $H$ itself measures the frequency of each element in the distribution adjusted for the relative size of each element: that is, $H$ is effectively a probability. But since to each probability, one can associate an information measure by taking the log-probability, we have an information interpretation of $H$ in terms of

$$R_2(Q(X)) = -\log H(X), \tag{17}$$

namely the second order Rényi entropy of the "rich get richer" probability system associated to $X$. The Rényi entropy [25] of a probability system $P = [p_1, ..., p_n]$ is more generally defined as

$$R_\lambda(P) = \frac{1}{1-\lambda} \log(\sum_{i=1}^{N} p_i^\lambda) \qquad (\lambda \geq 0, \lambda \neq 1). \tag{18}$$

The $I$ index is also a normalized version of the second order generalized inequality entropy measure, the general class of which plays a central role in the axiomatic approach to inequality measurement discussed shortly:

$$E_2 = \frac{1}{2}[\frac{1}{N}\sum_{i=1}^{N}(\frac{x_i}{\text{mean}(X)})^2 - 1] = \frac{1}{2}[NH - 1]. \tag{19}$$

In other words

$$I = \frac{2E_2}{2E_2 + 1}. \tag{20}$$

Finally, the new index has this other general interpretation also inherited from the $H$ index: usually this latter emerges as a measure of how much of the variability of an aggregate variable (e.g. GDP [19]) is due to the typical fluctuation of the individual components (firms' sales), provided these latter are statistically independent from one another. That is, the following interpretation holds in many contexts:

$$H = \frac{\text{aggregate volatility}}{\text{individual volatility}}. \tag{21}$$

In terms of the new inequality index, we have

$$\frac{\text{aggregate volatility}}{\text{individual volatility}} = \frac{1}{N}\frac{1}{1-I}. \tag{22}$$

Thus, a large aggregate volatility can emerge even in a very large system of independent components if these latter are highly unequal: $1/N$ is the damping factor due to the aggregation of independent homogenous components; but this damping factor is offset if the individual components are highly unequal. The argument in terms of the $H$ index is known [19].[16] But the reformulation in terms of the new index uncovers an

$$\text{Inequality Multiplier} = \frac{1}{1-I} = \sum_{k=0}^{\infty} I^k. \tag{23}$$

Any situation where the $H$ index plays a role, so could also do the inequality multiplier; and this is potentially the case of any sum of individual variables. Consider, for example, the return of an index or portfolio of assets, which is the average return of the individual assets, say $R = \sum_{i=1}^{N} x_i R_i$, where $[x_i] = X$ are the asset weights. In the simple case of independent, identically distributed, zero-mean individual asset returns, the portfolio's volatility is $\text{var}(R|X) = H(X)\text{var}(R_1)$.[17]

Relaxing the assumption of independent individual components, one can also define a linkage multiplier, the part of aggregate fluctuation due to the network amplification or propagation of interdependent individual shocks [28], or even consider a multiplier due to the inequality of network degrees (since some individuals are more heavily connected than others), usually measured by the coefficient of variation (*CV*) of the degrees [28]. The link between *CV* and *I* is a direct one:

---

[16] According to [19], the micro shocks of 100 largest firms in the US could account for one-third of aggregate fluctuations of the country's GDP.

[17] The $H$ index appears in portfolio theory as an effective portfolio diversification measure [26, 27].

$$I = \frac{CV^2}{CV^2 + 1}. \tag{24}$$

To summarize: the new index has potentially a range of applications due to its simple connections with a few core concepts across the sciences, connections given by the conversion formulas (already proven or easy to establish):

$$NH = CV^2 + 1 = (1-I)^{-1} = \sum_{k=0}^{\infty} I^k, \tag{25}$$

$$CV^2 = I(1-I)^{-1} = \sum_{k=1}^{\infty} I^k. \tag{26}$$

$$R_2 = \log[N(1-I)]. \tag{27}$$

$$E_2 = \frac{1}{2} \frac{I}{1-I}. \tag{28}$$

As an inequality measure, more specifically, the $I$ index enjoys the axioms of the normative approach to inequality measurement, which requires that an inequality measure $J = J(X)$ satisfy a list of properties, notably:

1. Normalization: $0 \leq J(X) \leq 1$.
2. Scale invariance: $J(\lambda X) = J(X)$ for $\lambda > 0$.
3. Anonymity: $J(X)$ should be invariant to the way the elements of $X$ are indexed (invariance to permutations of $X$).
4. Transfer principle: a rank-preserving transfer from a greater to lower element of $X$ (from a richer to a poorer person) should reduce inequality.
5. Population principle: $J(X, X) = J(X)$ (invariance to pooled replications of $X$).
6. Decomposability: if $X$ comes in subgroups, then $J$ should be decomposable into inequality within subgroups and inequality between subgroups.

We know of no inequality measure in the literature that satisfy all the 6 axioms. The Gini index satisfies Axioms 1-5, but not necessarily Axiom 6 (decomposability), except in the case of nonoverlapping subgroups [29, Appendix A]. By an important theorem [30-32], an inequality measure $J(X)$, continuous with respect to each element of $X$, satisfies the principle of transfers, scale independence, and decomposability if and only if it is some increasing function of a generalized entropy measure (where $\theta$ can be any real number including $\theta = 1$ if interpreted as $\theta \to 1$) [18]

$$E_\theta = \frac{1}{\theta(\theta-1)}[\frac{1}{N}\sum_{i=1}^{N}(\frac{x_i}{\text{mean}(X)})^\theta - 1]. \tag{29}$$

**Proposition 2**. *The index I satisfies the Axioms 1-6 listed above.*

Proof. The index $I$ obeys the population principle and anonymity in a straightforward way, and it is normalized. Since $I = 2E_2/(2E_2 + 1)$, a strictly increasing function of $E_2$, it satisfies the principle of transfers, scale independence, and decomposability by virtue of the above-mentioned theorem. ∎

One can also derive the 6 properties for $I$ by direct calculation; for example, the transfer principle: a rank-preserving transfer $t > 0$ from richer to poorer, or $(x_i, x_j) \to (x_i - t, x_j + t)$, is by definition one such that $x_i - t > x_j + t$, hence it is one such that $2t(x_j - x_i) < -4t^2$. The transfer preserves mean($X$) but decreases the second moment: mean($X^2$) $\to$ mean($X^2$) + $2t^2 + 2t(x_j - x_i)$. Thus, the transfer lowers $I$, namely $1 - (\text{mean}(X))^2/\text{mean}(X^2)$. The decomposition of $I$ by subgroup components can be established from the corresponding decomposition formula of the generalized entropy $E_2$ through a standard (if tedious) procedure [29, Appendix A].

## 4 Comparing the Two Inequality Measures

We emphasized at least two reasons in favor of the new index over the Gini index (that $I$ reflects better extreme variability and that it is always decomposable). A systematic empirical investigation of the two inequality measures using income or wealth distributions (in view notably of the first argued limitation of the Gini index) is beyond this paper's scope: for such study to be conclusive one would need to work with complete income or wealth distributions, rather than the relatively aggregated and often partial data, coming, say, by income classes, such as the US census data used in Figure 3. (We should not expect a major diffence between the two indices using aggregated data that even out extremes, although we note a turning point around 1995 when $G$ becomes greater than $I$.)[19] Thus we will be contempt here with theoretical comparisons of the two indices illustrated, when needed, with simulated heavy-tailed data.

---

[18] This is a central theorem of the axiomatic approach [33: see Theorem 4]. See also [34].

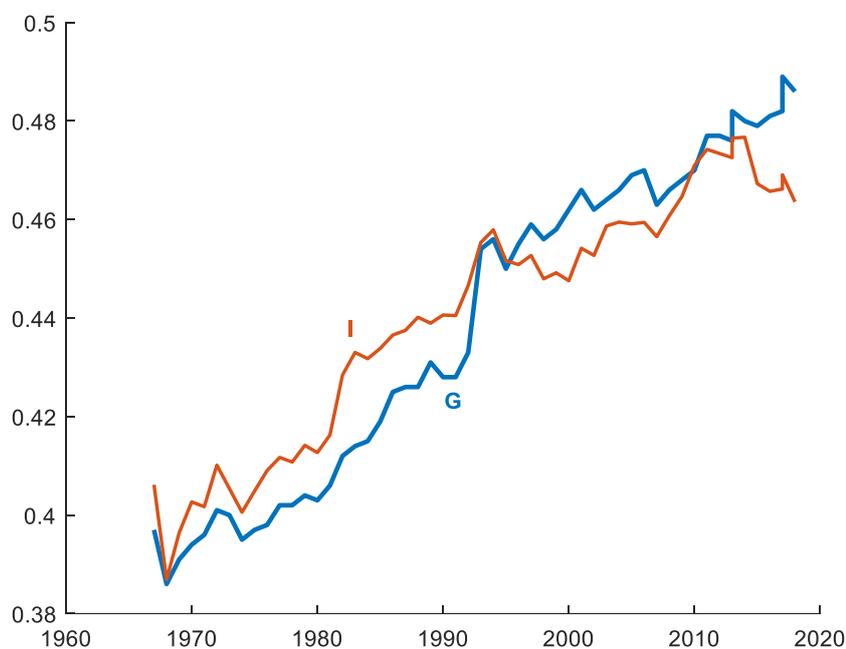

Figure 3. The inequality measures $G$ and $I$ for the USA census income data.

The two measures $I$ and $G$ share a few similarities. They coincide for at least two distributions: the exponential one, for which $I=G=0.5$, and the Bernoulli one, or the indicator function of an event $E$, for which $I = G = 1 - \text{prob}(E)$. For a Pareto with exponent $\alpha$, we have $G = G(\alpha) = (2\alpha - 1)^{-1}$ if $\alpha > 1$, as already mentioned earlier; in contrast $I = I(\alpha) = (\alpha - 1)^{-2}$ for $\alpha > 2$, as one can establish using the moment formula for the Pareto mentioned in the brief reminder on power laws (Section 2). Simple calculations yield $I(\alpha) \geq G(\alpha)$ if $2 < \alpha < 2 + \sqrt{2}$ and $I(\alpha) \leq G(\alpha)$ if $\alpha \geq 2 + \sqrt{2}$.

---

[19] Source: Historical Income Tables, census.gov. These are income data (in dollars) grouped in classes: [under 15000], [25,000 to 34,999], …, [200000 and over]. We take the lowest income to be zero and roughly estimate the highest income in such a way as to recover the summary statistics (such as the mean income) reported in the data. An overview of the limitations of common inequality data can be found e.g. in the recent review paper [35].

Conceptually, moreover, both indices are just normalized dispersion measures. In fact, the two indices belong to a general class of inequality measures:

$$I_p(X) = \frac{\frac{1}{2}\mathbb{E}|X - X'|^p}{\mathbb{E}|X|^p}, \tag{30}$$

where, again, $X'$ is an independent copy of $X$, and $p \geq 1$.[20] The general index $I_p$ is well-defined for any nonzero variable $X$ with $\mathbb{E}|X|^p < \infty$, and it is in the range [0,1]. This is true by the triangle inequality:[21]

$$(\mathbb{E}|X - X'|^p)^{1/p} \leq (\mathbb{E}|X|^p)^{1/p} + (\mathbb{E}|X'|^p)^{1/p} = 2(\mathbb{E}|X|^p)^{1/p}. \tag{31}$$

For $p = 2$, the numerator in (30) is $\mathbb{E}(X - X')^2/2 = \mathbb{E}(X^2 - 2X'X + X^2)/2 = \text{var}(X)$, so $I_2 = I$. For $p = 1$ we recover $G$ if $X \geq 0$ and $\mathbb{E}(X) > 0$. In fact we should define the Gini index more generally as $G(X) = I_1(X)$, namely

$$G(X) = \frac{\frac{1}{2}\mathbb{E}|X - X'|}{\mathbb{E}|X|}. \tag{32}$$

As announced earlier (Section 2), this general definition circumvents the undesirable behavior of the common definition of $G$ vis-à-vis signed variables. In particular, if $X$ is a symmetric variable (that is, $X$ and $-X$ have the same distribution), then $G(X) = G(-X)$ by the extended definition (32). However, one can show that $G = 1/\sqrt{2} \approx 0.7$ for the Gaussian, suggesting this latter is more unequal distribution than an exponential variable, for example; more generally, a significant presence of negatives in a distribution tends to produce relatively high values for both indices, which is perhaps intuitively reasonable: a society with symmetric income distribution would be one in which only half of the population have positive incomes and the other half are indebted to them, an extreme form of inequality! This is even better reflected in the new index, since $I(X)=1$ if $\text{mean}(X)=0$. But then both indices (especially the new one) are poor measures of heavy-tailedness for signed variables (by ranking higher a Gaussian compared to say a Pareto with index 1.5, for example: the next paragraphs define $I$ for infinite-variance variables). This is easily fixed however: the heavy-tailedness of $X$ can be measured by $I(|X|)$.

---

[20] If we define dispersion more strictly in terms of the distance function $(X, Y) \mapsto \|X - Y\|_p$ induced by the norm $X \mapsto \|X\|_p = (\mathbb{E}|X|^p)^{1/p}$, then we would consider instead $I^{1/p}$ as inequality measure. But statisticians use more flexible "distance" concepts (such as a square distance).

[21] The triangle inequality is applied, that is, to the distance function referred to in the preceding note.

We close the comparison of the two indices by a discussion of their respective statistical stability. One might be inclined to prejudge that $I$ is a more volatile statistic compared to $G$, and to assume that $I$ is the more volatile, the heavier the tail of $X$. But this is only true of the ratios' numerators. In fact, $I$ is more stable statistically then $G$ for infinite-variance variables, paradoxically: moreover, we have the curious pattern that $I(X)$ is the more stable compared to $G(X)$, the more volatile is $X$; and the reverse pattern for $G$ (Figure 4).

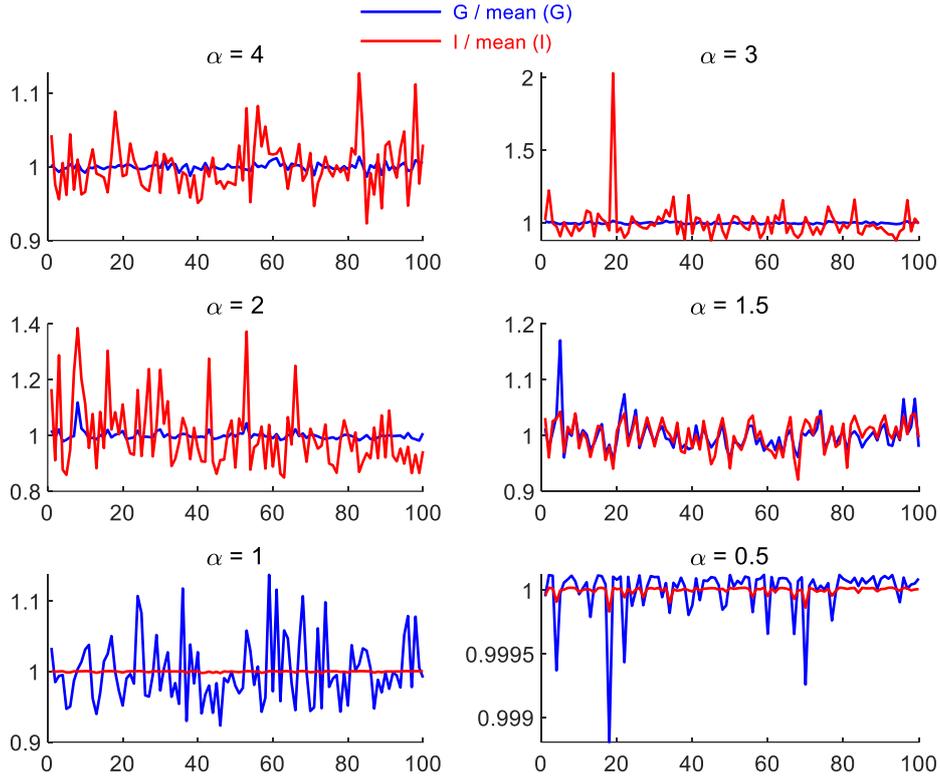

Figure 4. Statistical Stability of *G* versus *I* for Pareto variables (100 large samples of 100000 random draws). Notice the curious pattern: the lower is alpha (hence the heavier-tailed the Pareto), the more stable is *I* (red) compared to *G* (blue), and vice versa.

The slow convergence pattern of *G* is not surprising per se and it has been documented [8]. What is surprising is the inverse pattern of *I* compared to that of *G*. This suggests a scrutiny of the limit behavior of the *I* index. Formally, we model the possible sample realization of *I(X)* by the following estimator, based on *N* independent copies $\{X_1, ..., X_N\}$ of *X*, to investigate its large-sample behavior:

$$I_N(X) = 1 - \frac{\sum_{i=1}^{N} X_i^2}{N(\sum_{i=1}^{N} X_i)^2}. \tag{33}$$

Because $I_N(X)$ involves sums of independent copies of *X* and $X^2$, its limit behavior is regulated by the general central limit theorem (in the same way as for the *H* index, whose limit behavior for a Pareto with $\alpha \geq 1$ is known [19, prop. 2]).

***Proposition 3.*** *If X is a positive power law with exponent $\alpha \in (0,2)$, then $I_N(X) \to 1$ as $N \to \infty$. Moreover, $1 - I_N(X) \to 0$ like $N^{-1}$ if $\alpha \in (0,1)$, $N^{-1+2/\alpha}$ if $\alpha \in (1,2)$, and $N^{-1}(\log N)^2$ if $\alpha = 1$ and X is Pareto.*[22]

This proposition, proven in the Appendix, explains the above-mentioned seemingly paradoxical statistical stability of the index *I* (Figure 4): the heavier the tail of the power law, the more stable the index *I*, with the possible exception of $\alpha = 1$ (which more generally is a singular parameter in the theory of the limit sum of power laws).

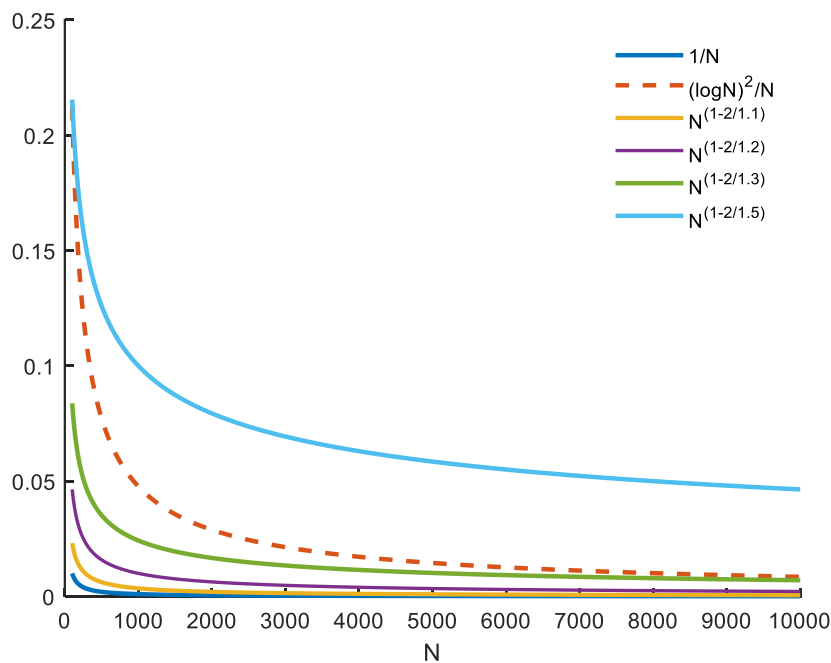

Figure 5. Convergence rate of the empirical (or sample) inequality index $I_N(X)$ for infinite-variance Pareto variable *X*. The lower is the tail exponent $\alpha$, the more rapid the convergence rate (for $\alpha \neq 1$).

Proposition 3 suggests that we extend the definition of the *I* index to a theoretical variable *X* as the large sample limit of its empirical version, namely as $I(X) = \lim_{N \to \infty} I_N(X)$, which for a finite-variance distribution reduces of course to $\mathrm{var}(X)/\mathbb{E}(X^2)$, by the law of large numbers, and for an (infinite-variance) power law $X$ with $\alpha \in (0,2)$ is $I(X) = 1$ (by Proposition 3). An apparent limitation of this extension to infinite-variance theoretical distributions is that it does not reflect the gradation of inequality for power laws of exponent in the range $\alpha \in (0,2)$, a scale of inequality that we know is inversely related to $\alpha$. But this is hardly a real problem, for the gradation of inequality for power laws with $\alpha \in (0,2)$ is recovered empirically in the speed of the convergence $I_N(X) \to 1$, which is the more rapid, the lower is $\alpha$, so that for a comparably large sample size $N$, the index $I_N(X)$ tends to be higher, the lower is $\alpha$.

---

[22] We assume X≥0 merely to avoid worrying about an additional condition: see Appendix.

The limit behavior of the Gini index, on the other hand, is that of the estimator

$$G_N(X) = 2 \frac{\sum_i^N i X_{(i)}}{N \sum_i^N X_i} - 1. \tag{34}$$

The study of $G_N(X)$ involves more care and technique because $G_N(X)$ involves a sum of ranked (hence dependent) variables in the numerator. For a power law $X$, a characterization of the limit behavior of $G_N(X)$ has already be done for the case $1 < \alpha < 2$ [8], which show the slow convergence of $G_N(X)$ for infinite-variance power laws: the complement case $0 < \alpha < 1$ would involve essentially the same technicalities, which we do not repeat here. We mention however:

**Proposition 4.** *If $X$ is a power law with exponent $\alpha < 1$, then $G_N(X) \to 1$ as $N \to \infty$.*

In lieu of a formal but long proof, we just emphasize here a simple intuitive argument as to why this result is to be expected (referring the reader to [8] for the preliminary technique needed before one can apply the general central limit theorem for a formal proof). The intuitive argument is based on Atkinson's formula [36], according to which "when a very top group of the income distribution, infinitesimal in numbers, owns a finite share $S$ of total income, the Gini coefficient $G$ can be approximated by $G^*(1-S) + S$, where $G^*$ is the Gini coefficient for the rest of the population" [37]. In a large sample from a power law with $\alpha < 1$, the highest

"income" on average has the share $S = 1 - \alpha$ according to equation (12), and we also have $G = G^*$ approximately (because $G$ is only mildly affected if one observation, here the biggest one, is removed). So $G = \alpha G + (1 - \alpha)$, hence $G \equiv \lim_{N \to \infty} G_N = 1$.

## 5 Summary

Beyond the specific discussion on inequality measurement, this paper revisits received doctrine on robustness, whose desirability might be questionable for heavy tailed variables. The Gini index remains a powerful inequality measure for light-tailed distributions potentially distorted by outliers. But the variance normalized by the second moment better suits heavy-tailed variables, and it satisfies the normative axioms of inequality measurement, including decomposability. As a concept, the new index is potentially useful beyond the specific purpose of inequality measurement due to its simple relationships with fundamental concepts across the sciences, such as: variance, Herfindahl-Hirschman index, and Rényi entropy.

## Appendix

For two functions $F$ and $G$ of $x$, the notation $F \sim G$ means $F(x)/G(x) \to 1$ as $x \to \infty$. For variables $\{X_N\}$ and $Y$, $X_N \sim a_N + b_N Y$ means the convergence in distribution $(X_N - a_N)/b_N \to Y$ holds. Many claims in the main text rely on the following [15: sec. 2.2, th. 2.2.15, 16: sec. 3.8, th. 3.8.2]:

***Limit Theorem for Power Laws.*** Let $\{X_1,...,X_N\}$ be independent copies of a variable $X$ with $\text{prob}\{|X| \geq x\} \sim Cx^{-\alpha}$, $C > 0$, $\alpha \in (0,2)$, and $\lim_{x \to \infty}[\text{prob}\{|X| \geq x\}/\text{prob}\{X \geq x\}] = \theta \in [0,1]$. Then $N^{-1/\alpha}(\sum_{i=1}^{N} X_i - b_N) \to \Lambda(\alpha)$ as $N \to \infty$, where $\Lambda(\alpha)$ is a Levy variable with exponent $\alpha$, and $b_N = 0$ if $\alpha \in (0,1)$, $N\mathbb{E}(X)$ if $\alpha \in (1,2)$, and $N\mathbb{E}(X 1_{|X| \leq N})$ if $\alpha = 1$.

For a Pareto $X \geq 1$, the theorem says $N^{-1/\alpha}(\sum_{i=1}^{N} X_i - b_N) \to \Lambda(\alpha)$ as $N \to \infty$, where

$$b_N = \begin{cases} 0, & \alpha < 1 \\ N \log N, & \alpha = 1 \\ E(X)N, & \alpha > 1. \end{cases} \quad (35)$$

***Proof of Proposition 3.*** Without loss of generality, assume $X \geq 0$ (otherwise, simply add the condition $\theta \in [0,1]$). We investigate the limit behavior of

$$I_N(X) = 1 - \frac{(\sum_{i=1}^{N} X_i)^2}{N \sum_{i=1}^{N} X_i^2}.$$

By the limit theorem, $\sum_{i=1}^{N} x_i \sim b_N(\alpha) + N^{1/\alpha}\Lambda(\alpha)$ and $\sum_{i=1}^{N} x_i^2 \sim N^{2/\alpha}\Lambda(\alpha/2)$.[23] Thus

$$I_N(X) \sim 1 - \frac{[b_N(\alpha) + N^{1/\alpha}\Lambda(\alpha)]^2}{N^{1+2/\alpha}\Lambda(\alpha/2)}. \quad (36)$$

Hence

$$1 - I_N(X) \sim \begin{cases} \dfrac{[\Lambda(\alpha)]^2}{\Lambda(\alpha/2)} \dfrac{1}{N}, & \alpha \in (0,1) \\ \dfrac{[\mathbb{E}(X)]^2}{\Lambda(\alpha/2)} N^{1-2/\alpha}, & \alpha \in (1,2) \end{cases} \quad (37)$$

---

[23] The second sum is just an application of the theorem to $X^2$ which is a Pareto with exponent $\alpha/2$, since $\text{prob}\{X^2 \geq z\} = \text{prob}\{X \geq z^{1/2}\} = (z^{1/2})^{-\alpha}$.

In particular, $I_N(X) \to 1$ as $N \to \infty$ in either case. The case $\alpha = 1$ requires special care; for a Pareto, we have $1 - I_N(\alpha) \sim C_1 N^{-1} [\log N]^2 / \Lambda(1/2) \to 0$. More generally, $b_N(1) = N\mathbb{E}(X 1_{X \leq N}) = N \int_0^N L(x) x^{-1} dx$, where by assumption $L(x) \to C$ as $x \to \infty$, a slowly-varying function, which therefore obeys $[L(N)]^{-1} \int_0^N L(x) x^{-1} dx \to \infty$ as $N \to \infty$ [11, Remark 1.2.7], hence $\int_0^N L(x) x^{-1} dx \to \infty$ [since $L(N) \to C$], that is, $b_N(1)/N \to \infty$. This suggests writing (36) for $\alpha = 1$ as follows:

$$[1 - I_N(X)]^{1/2} \sim \frac{[b_N(1) + N\Lambda(1)]}{N^{3/2}[\Lambda(1/2)]^{1/2}} = \frac{1 + [N/b_N(1)]\Lambda(1)}{[b_N(1)/N]N^{5/2}[\Lambda(1/2)]^{1/2}} \to 0. \tag{38}$$

All in all, $I_N(X) \to 1$ for every $\alpha \in (0, 2)$. ∎